\documentclass[12pt]{article}
\usepackage{graphicx}
\usepackage[cp1251]{inputenc}
\usepackage{rotating}
\begin{document}
 \noindent {\footnotesize\it Astronomy Reports, 2020, Vol. 64, No 12, pp. 1026--1033}
 \newcommand{\dif}{\textrm{d}}

 \noindent
 \begin{tabular}{llllllllllllllllllllllllllllllllllllllllllllll}
 & & & & & & & & & & & & & & & & & & & & & & & & & & & & & & & & & & & & & \\\hline\hline
 \end{tabular}

 \vskip 0.5cm
  \centerline{\bf\large Kinematic Properties of Young Intermediate- and}
  \centerline{\bf\large Low-Mass Stars from the Gaia DR2 Catalogue}
 \bigskip
 \bigskip
  \centerline
 {
 V.V. Bobylev and A.T. Bajkova
 }
 \bigskip
 \centerline{\small \it
 Central (Pulkovo) Astronomical Observatory, Russian Academy of Sciences,}
 \centerline{\small \it Pulkovskoe shosse 65, St. Petersburg, 196140 Russia}
 \bigskip
 \bigskip
 \bigskip

 {
{\bf Abstract}---We have studied the kinematic properties of young pre-main-sequence stars. We have selected
these stars based on data from the Gaia DR2 catalogue by invoking a number of photometric infrared surveys.
Using 4564 stars with parallax errors less than 20\%, we have found the following parameters of the angular
velocity of Galactic rotation:
$\Omega_0 =28.84\pm0.10$~km s$^{-1}$ kpc$^{-1}$,
 $\Omega^{'}_0=-4.063\pm0.029$~km s$^{-1}$ kpc$^{-2}$ and
 $\Omega^{''}_0=0.766\pm0.020$~km s$^{-1}$ kpc$^{-3}$, where the Oort constants are
$A= 16.25\pm0.33$~km s$^{-1}$ kpc$^{-1}$ and $B=-12.58\pm0.34$~km s$^{-1}$ kpc$^{-1}$.
The circular rotation velocity of the solar neighborhood around the Galactic
center is $V_0=230.7\pm4.4$~km s$^{-1}$ for the adopted Galactocentric distance of the Sun $R_0=8.0\pm0.15$~kpc. The residual velocity dispersion for the stars considered is shown to be low, suggesting that they are extremely young. The residual velocity dispersion averaged over three coordinates is $\sim$11 km s$^{-1}$ for Herbig Ae/Be stars and $\sim$7 km s$^{-1}$ for T Tauri stars.
  }

\medskip DOI: 10.1134/S1063772920120033

 \section{INTRODUCTION}
Among pre-main-sequence stars there are both intermediate-mass stars ($2M_\odot-8M_\odot$), i.e., Herbig
Ae/Be stars, and less massive T Tauri stars with masses $<2M_\odot$. Such stars are of great interest for studying the structure and kinematics of the Galactic disk owing to their being exceptionally young. However, the astrometric data needed for an analysis have been known quite recently only for a few hundred such stars located in the Gould Belt, i.e., near the Sun.

In this regard the situation changed abruptly with the appearance of the Gaia DR2 catalogue [1, 2],
which contains the trigonometric parallaxes and proper motions of $\sim$1.3 billion stars. For a relatively
small fraction of these stars their line-of-sight velocities have been measured. In the Gaia catalogue [3] the
photometric measurements are presented in two wide
bands and, therefore, only a very rough classification
of stars is possible. For a reliable classification it is
necessary to invoke more accurate spectroscopic and
photometric data from other sources. A number of important studies related to the kinematics of various
Galactic subsystems have been performed on the basis of Gaia DR2 data.

At present, several samples of young pre-mainsequence stars from the Gaia DR2 catalogue are
known. For example, based on kinematic and photometric
data, Zari et al. [4] selected more than 40000 T Tauri stars from the Gaia DR2 catalogue. All these
stars are no farther than 500 pc from the Sun and are
closely associated with the Gould Belt. Their spatial and kinematic properties were studied in detail in [5].

More than a million candidates for young stellar objects located in active star-forming regions were
identified in [6]. For this purpose, the Gaia DR2 data were combined with the infrared 2MASS [7] and
WISE [8, 9] photometry by invoking the data on interstellar
extinction from the Planck experiment [10].
Four classes of objects were considered: young stellar
objects, extragalactic objects, main-sequence stars,
and evolved stars. More than 25 000 objects are low-mass
pre-main-sequence stars. Their kinematic properties were analyzed in [11], where the Galactic rotation
parameters were estimated.

A technique for the selection of pre-mainsequence stars from the Gaia DR2 catalogue different
from the one in [6] was proposed in [12]. Here the photometric measurements from the Gaia DR2 catalogue
are used in combination with the infrared
2MASS, WISE, IPHAS [13, 14], and VPHAS+ [15]
data. Among 8000 candidates the astrometric measurements
have a high accuracy (the parallax error is
mostly small) only for half of them. Furthermore,
these authors found that the fraction of stars common to the sample from [6] does not exceed 50\%.

The goal of this paper is to refine the Galactic rotation parameters based on the sample of young stars
from [12] and to determine the parameters of the residual velocity ellipsoid for these stars. Only multiband
photometry was used in [12] for the selection of stars and, therefore, confirming the youth of these
stars based on a kinematic analysis is a topical problem.

 \section{DATA}
The selection of young stars in [12] was made using machine learning techniques based on data from the
Gaia DR2 catalogue by invoking infrared 2MASS, WISE, IPHAS, and VPHAS+ photometric measurements.
A total of 4~150~983 objects were identified by them. In this paper we use the stars from the PMS,
CBe, and EITHER samples whose members have a high probability of belonging to them.

Note that an analysis of 48 photometric differences provides a basis for the classification selection of stars
in [12]. They were calculated using data from the Gaia DR2 catalogue, where the photometric G$_{BP}$, G and G$_{RP}$ bands have mean wavelengths of 0.50, 0.59, and 0.77~$\mu$m, respectively, the and bands from the
IPHAS and VPHAS+ catalogues have 0.62 and 0.66~$\mu$m, respectively, the J, H, and K$_s$ bands from
the 2MASS catalogue have 1.24, 1.66, and 2.16~$\mu$m, respectively, and the W1, W2, W3 and W4 bands
from the WISE catalogue have 3.4, 4.6, 12, and 22~$\mu$m, respectively. Therefore, when selecting the initial data, these authors gave great attention to the quality of the photometric data. As a result, for a number of stars included in the working sample of 4~150~983 stars there are those for which the Gaia DR2 catalogue contains no measurements of their parallaxes. Furthermore, the final sample has the limitation in Galactic latitude $|b|<5.5^\circ$ due to such limitations in the IPHAS and VPHAS+ catalogues.

Talking about the completeness, these authors have in mind the completeness of the selection of stars
of a certain type (Herbig Ae/Be, classical Be, or PMS  stars) from the working data set (from 4~150~983 stars).
Based on an analysis of the photometric differences, they assigned the probability $p$ of belonging to each of
the types considered (Herbig Ae/Be, classical Be, or PMS stars) to each star.

The PMS sample consists of 8470 candidates for young pre-main-sequence stars. The completeness of
this sample was estimated in [12] to be 78.8\%, and its
members are both Herbig Ae/Be and T Tauri stars.
Out of the above three samples, this is the richest in
stars and, therefore, in this paper it is the main one for our kinematic analysis.

The CBe sample contains 693 candidates for classical Be stars. Its completeness is 85.5\%. CBe are stars
of spectral type B on the main sequence with rapid axial rotation and surrounded by gas disks. These are
the most massive stars among those considered by us.

The EITHER sample includes 1309 stars. It is composed of stars with the following probabilities $p$: $(p_{\rm PMS}+p_{\rm CBe})>50\%,$ but $p_{\rm PMS}<50\%$ and $p_{\rm CBe}<50\%$.

\begin{figure}[t]
{\begin{center}
   \includegraphics[width=0.7\textwidth]{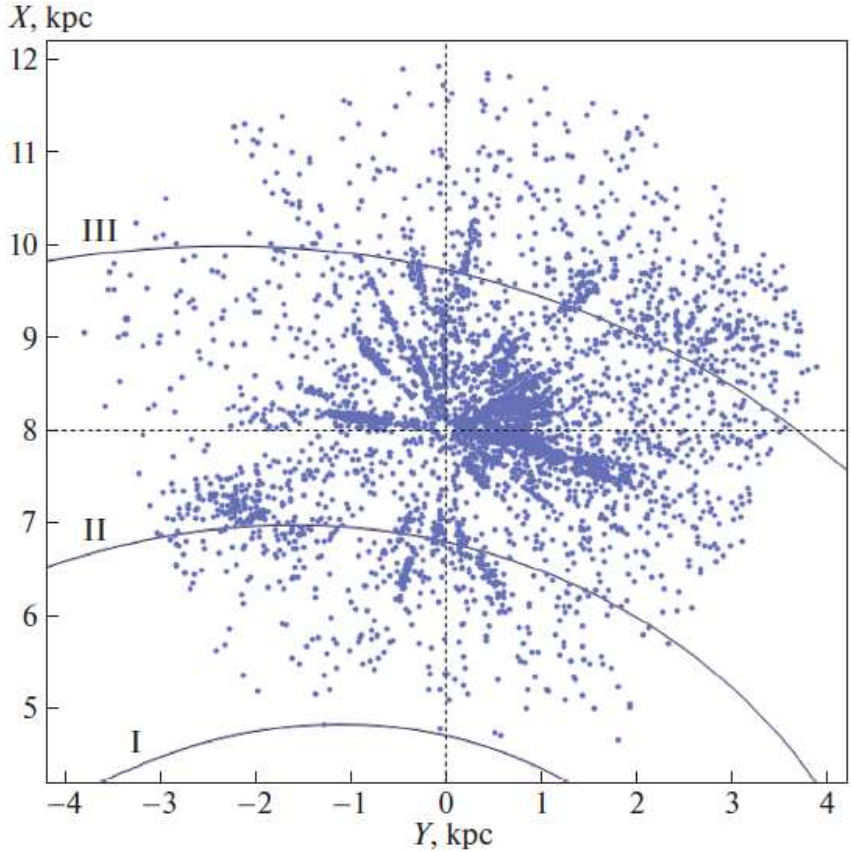}
 \caption{
The distribution of stars from the ALL sample with relative trigonometric parallax errors less than 20\% on the Galactic $XY$ plane; the spiral pattern with a pitch angle of $-13^\circ$ from [16] is presented.
  } \label{f-1}
\end{center}}
\end{figure}

All of the listed samples contain very young stars. There are no overlaps between the samples. To use a
maximum number of young stars in determining the Galactic rotation parameters, we also produced the
combined ALL sample composed of the stars from all three CBe, EITHER, and PMS samples. The distribution
of stars from this sample with relative trigonometric parallax errors less than 20\% on the Galactic $XY$
plane is shown in Fig. 1. A connection with the
spiral structure can be seen in the figure. First, the segment
of the Local arm (Orion arm) passing near the
Sun at an angle of about $-15^\circ$ to the $Y$ axis is clearly
seen; second, a concentration of stars is seen near the
Perseus arm (designated as III) in the second Galactic
quadrant; third, there are two clumps of stars in the
region of the Carina–Sagittarius arm segment (II).
Interestingly, only the Local arm segment is clearly
seen in the distribution of the candidates for T Tauri stars [11, Fig. 3] selected in [6] from the spiral structure.

 \section{METHODS}
 \subsection{Galactic Rotation Parameters}
We use a rectangular coordinate system centered on the Sun, in which the $x$ axis is directed toward the Galactic center, the $y$~axis is toward the Galactic rotation, and the $z$~axis is toward the north pole of the Galaxy. Then $x=r\cos l\cos b,$ $y=r\sin l\cos b$ and $z=r\sin b.$
From observations there are known three components of a star velocities: the line-of sight velocity $V_r$ and two projections of the tangential velocity $V_l=4.74r \mu_l\cos b$ and $V_b=4.74r\mu_b$, directed along the Galactic longitude $l$ and latitude $b$ respectively. All the velocity components are measured in km s$^{-1}$. Here, the coefficient 4.74 is the ratio of the number of kilometers in an astronomical unit to the number of seconds in a tropical year, and $r=\pi^{-1}$~is the heliocentric distance of the star in kpc, which we calculate through the parallax of the star $\pi$ in mas. The components of the proper motion $\mu_l\cos b$ and $\mu_b$ are expressed in mas year$^{-1}$.

To determine the parameters of the Galactic rotation curve, we use equations obtained from the Bottlinger formulas, in which the angular velocity $\Omega$ is expanded into a Taylor series in powers of $(R-R_0)$ to terms of the second order of smallness $r/R_0$. There are very few stars with measured
line-of-sight velocities in our samples. Therefore, for our analysis we use only the following two equations
with the proper motions on the left-hand sides:
 \begin{equation}
 \begin{array}{lll}
 V_l= U_\odot\sin l-V_\odot\cos l-r\Omega_0\cos b\\
 +(R-R_0)(R_0\cos l-r\cos b)\Omega'_0
 +0.5(R-R_0)^2(R_0\cos l-r\cos b)\Omega''_0,
 \label{EQ-2}
 \end{array}
 \end{equation}
 \begin{equation}
 \begin{array}{lll}
 V_b=U_\odot\cos l\sin b + V_\odot\sin l \sin b-W_\odot\cos b\\
    -R_0(R-R_0)\sin l\sin b\Omega'_0
    -0.5R_0(R-R_0)^2\sin l\sin b\Omega''_0,
 \label{EQ-3}
 \end{array}
 \end{equation}
where $R$~is the distance from the star to the axis of Galactic rotation (cylindrical radius):
 \begin{equation}
 R^2=r^2\cos^2 b-2R_0 r\cos b\cos l+R^2_0.
 \end{equation}
The quantity $\Omega_0$ is the angular velocity of the Galaxy at a solar distance $R_0$, $\Omega_0^{(i)}$~is the $i$-th  derivative of the angular velocity with respect to $R$, the linear rotation velocity at a solar distance equals to $V_0=R_0\Omega_0$, $R_0$~is the galactocentric distance of the Sun.
In Eqs. (1) and (2) six unknowns are to be determined: $U_\odot,$ $V_\odot,$ $W_\odot,$ $\Omega_0,$ $\Omega^\prime_0$ and $\Omega^{\prime\prime}_0$. The Oort constants and are also of interest;
their values can be found from the expressions
\begin{equation}
 A=0.5\Omega^{\prime}_0R_0,\quad B=-\Omega_0+A, \label{AB}
\end{equation}
The kinematic parameters are determined by solving the conditional equations (1) and (2) by the least-squares
method. Weights of the form $w_l=S_0/\sqrt {S_0^2+\sigma^2_{V_l}}$ and
 $w_b=S_0/\sqrt {S_0^2+\sigma^2_{V_b}},$ where  $S_0$ is the ``cosmic'' dispersion,
$\sigma_{V_l}$ and $\sigma_{V_b}$ are the errors in the corresponding observed velocities, are used. $S_0$ is comparable to the root-mean-square residual $\sigma_0$ (the error per unit weight), which is calculated by solving the conditional equations (1) and (2). In this paper the adopted values of lie in the range 7--12 km s$^{-1}$. The system of equations (1) and (2) is solved in several iterations by applying the criterion to exclude the stars with large residuals.

 \subsection{Choosing $R_0$}
At present, a number of studies devoted to determining the mean Galactocentric distance of the Sun
using the individual determinations of this quantity in the last decade by independent methods have been
performed. For example,  $R_0=8.0\pm0.2$~kpc [17], $R_0=8.4\pm0.4$~kpc [18] or $R_0=8.0\pm0.15$~kpc [19].

Note also some of the first-class individual determinations of this quantity made in recent years. Based
on the masers from the Japanese VERA program, Hirota et al. [20] obtained an estimate of $R_0=7.9\pm0.3$~kpc. Having analyzed their 16-year-long series of observations of the motion of the star S2 around the supermassive black hole at the Galactic center, Abuter et al. [21] found $R_0=8.178\pm0.022$~kpc. Based on an independent analysis of the orbit of the star S2, Do et al. [22] found $R_0=7.946\pm0.032$~kpc.
Based on the listed results, in this paper we adopt $R_0=8.0\pm0.15$~kpc.

 \subsection{Calculating the Distances}
A problem with the Gaia DR2 trigonometric parallaxes has been known since the publication of the Gaia
DR2 catalogue: a correction with a value from 0.03 to 0.05 mas is needed [2, 23]. Taking into account the
determinations of this correction in [24--26], we must add a correction of 0.05 mas to all of the original stellar parallaxes from the Gaia DR2 catalogue, i.e., $\pi_{new}=\pi+0.05$~mas.

In this paper we usually calculate the distances via the trigonometric parallaxes. However, apart from the
Gaia DR2 parallaxes, the distances calculated in [27] by taking into account the peculiarities in the distribution of stars in the Galaxy are given in [12]. This
method is similar to the correction for the Lutz–Kelker bias [28], which can be taken into account at
large relative parallax errors. In this paper we use stars
with parallax errors less than 20\% and, therefore, these
corrections are negligible. For comparison, based on
the PMS sample, we derived the kinematic parameters using both systems of distances.

 \subsection{Residual Velocity Ellipsoid}
The stellar residual velocity dispersions are estimated using the following method [29]. Six second-order
moments $a, b, c, f, e, d$ are considered:
\begin{equation}
 \begin{array}{lll}
 a=\langle U^2\rangle-\langle U^2_\odot\rangle,\quad
 b=\langle V^2\rangle-\langle V^2_\odot\rangle,\quad
 c=\langle W^2\rangle-\langle W^2_\odot\rangle,\\
 f=\langle VW\rangle-\langle V_\odot W_\odot\rangle,~~
 e=\langle WU\rangle-\langle W_\odot U_\odot\rangle,~~
 d=\langle UV\rangle-\langle U_\odot V_\odot\rangle,
 \label{moments}
 \end{array}
 \end{equation}
which are the coefficients of the equation for the surface
 \begin{equation}
 ax^2+by^2+cz^2+2fyz+2ezx+2dxy=1,
 \end{equation}
and also the components of the symmetric tensor of moments of the residual velocities:
 \begin{equation}
 \left(\matrix {
  a& d & e\cr
  d& b & f\cr
  e& f & c\cr }\right).
 \label{ff-5}
 \end{equation}
In the absence of line-of-sight velocities, the following
three equations are used to determine the values of this
tensor:
\begin{equation}
 \begin{array}{lll}
 V^2_l= a\sin^2 l+b\cos^2 l\sin^2 l\\
 -2d\sin l\cos l,
 \label{EQsigm-2}
 \end{array}
 \end{equation}
\begin{equation}
 \begin{array}{lll}
 V^2_b= a\sin^2 b\cos^2 l+b\sin^2 b\sin^2 l+c\cos^2 b\\
 -2f\cos b\sin b\sin l
 -2e\cos b\sin b\cos l
 +2d\sin l\cos l\sin^2 b,
 \label{EQsigm-3}
 \end{array}
 \end{equation}
\begin{equation}
 \begin{array}{lll}
 V_lV_b= a\sin l\cos l\sin b+b\sin l\cos l\sin b\\
 +f\cos l\cos b-e\sin l\cos b
 +d(\sin^2 l\sin b-\cos^2\sin b),
 \label{EQsigm-4}
 \end{array}
 \end{equation}
which are solved by the least-squares method for the six unknowns $a,b,c,f,e,d$. The eigenvalues of the tensor (7) $\lambda_{1,2,3}$ are then found from the solution of the secular equation
 \begin{equation}
 \left|\matrix
 {
a-\lambda&          d&        e\cr
       d & b-\lambda &        f\cr
       e &          f&c-\lambda\cr
 }
 \right|=0.
 \label{ff-7}
 \end{equation}
The eigenvalues of this equation are equal to the reciprocals of the squares of the semiaxes of the velocity
moment ellipsoid and, at the same time, the squares of the semiaxes of the residual velocity ellipsoid:
 \begin{equation}
 \begin{array}{lll}
 \lambda_1=\sigma^2_1, \lambda_2=\sigma^2_2, \lambda_3=\sigma^2_3,\qquad
 \lambda_1>\lambda_2>\lambda_3.
 \end{array}
 \end{equation}
The directions of the principal axes of the tensors (11) $L_{1,2,3}$ and $B_{1,2,3}$ are found from the relations
 \begin{equation}
 \tan L_{1,2,3}={{ef-(c-\lambda)d}\over {(b-\lambda)(c-\lambda)-f^2}},
 \label{ff-41}
 \end{equation}
 \begin{equation}
 \tan B_{1,2,3}={{(b-\lambda)e-df}\over{f^2-(b-\lambda)(c-\lambda)}}\cos L_{1,2,3}.
 \label{ff-42}
 \end{equation}

\begin{figure}[t]
{\begin{center}
   \includegraphics[width=0.85\textwidth]{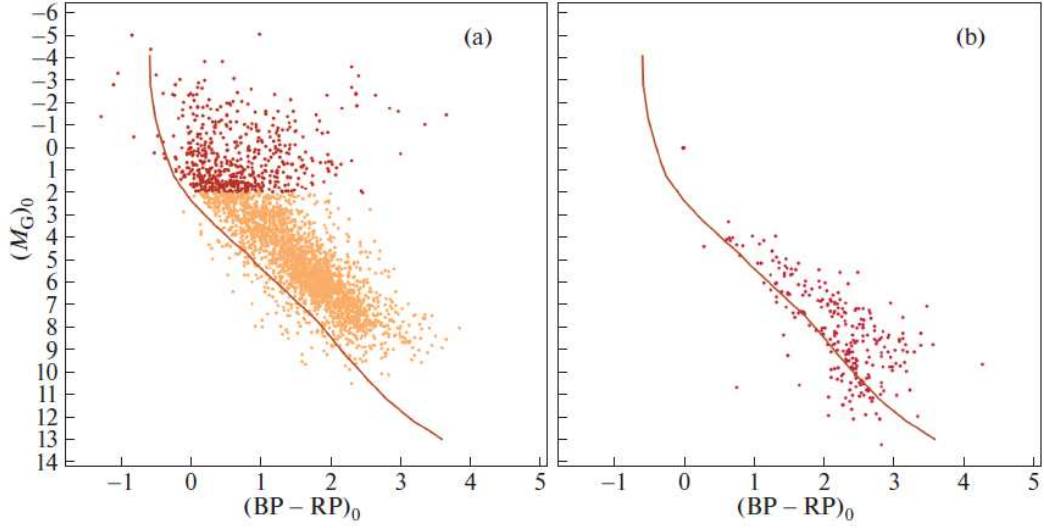}
 \caption{
The color–absolute magnitude diagram constructed from the stars of the PMS sample in the distance range $r=0.5-4$ kpc (a) and the nearest stars of this sample with $r\leq0.5$~kpc (b); the solid line indicates the main sequence.
   } \label{f-2}
\end{center}}
\end{figure}

 \section{RESULTS AND DISCUSSION}
The Galactic rotation parameters determined from the stars of the PMS sample are given in the upper part
of Table 1. For this purpose, we took the stars with relative trigonometric parallax errors less than 20\% and heliocentric distances less than 4 kpc. Here, the distances to the stars were calculated via the parallaxes
from the Gaia DR2 catalogue by adding the correction $\Delta\pi=0.050$~mas, then $r=1/(\pi+0.050)$~kpc. The
first, second, and third columns in the table give the solutions obtained from all stars of the PMS sample,
the stars from the upper part of the Hertzsprung–Russell diagram ($(M_G)_0\leq2^m$), and the stars from the
lower part of the Hertzsprung–Russell diagram ($(M_G)_0>2^m$), respectively. The results obtained by
excluding the nearest stars ($r>0.5$~kpc) are presented in the second and third columns of the table. Finally,
the results obtained from the nearest stars ($r\leq0.5$~0.5 kpc) are presented in the last column of the table.
We made the separation into parts with the boundary $(M_G)_0=2^m$ so as to have an approximately equal number
of stars in the samples. Note that such parameters as the linear rotation velocity of the solar neighborhood $V_0$ and the Oort constants $A$ and $B$ were estimated by taking into account the error in $R_0$ of
$\pm0.15$~kpc.

The parameters of the stellar residual velocity ellipsoid are given in the lower part of Table 1. When forming
the residual velocities of the stars, we made corrections for their group motion $(U,V,W)_\odot$ and the Galactic rotation.

The Hertzsprung–Russell (HR) diagram constructed from the stars of the PMS sample is shown in
Fig. 2. The corrections for extinction based on Gaia DR2 data have already been made in [12]. To construct
the zero-age main sequence, we used the evolutionary tracks from the PARSEC1 electronic database
[30] of version 1.2S [31, 32] with $A_V=0^m$ and solar metallicity ($Z=0.0152$).

The results presented in the last column of Table 1 and Fig. 2b show that the properties of the nearest
stars differ sharply from those of the more distant ones. They have a very high residual velocity dispersion,
i.e., may belong to stellar streams. A significant fraction of them lie on the main sequence, i.e., they
may be old. As a result, we concluded that it is better to exclude the nearest stars from consideration.

 \begin{table}[t]
 \caption[]{\small
The kinematic parameters found from the stars of the PMS sample with relative trigonometric parallax errors less than 20\%.
 }
  \begin{center}  \label{t:01}  \small
  \begin{tabular}{|l|r|r|r|r|r|}\hline
       Parameters & All stars &     $r>0.5$~kpc &     $r>0.5$~kpc & $r\leq0.5$~kpc \\
                  &  &  $(M_G)_0\leq2^m $ &   $(M_G)_0>2^m$ &                \\\hline

       $U_\odot,$ km s$^{-1}$ & $6.95\pm0.15$ & $5.83\pm0.35$ & $7.08\pm0.15$ &  $9.03\pm0.74$ \\
       $V_\odot,$ km s$^{-1}$ & $9.56\pm0.24$ & $9.05\pm0.56$ & $9.74\pm0.29$ & $11.02\pm1.33$ \\
       $W_\odot,$ km s$^{-1}$ & $7.47\pm0.12$ & $7.21\pm0.26$ & $7.92\pm0.12$ &  $5.60\pm0.48$ \\
  $\Omega_0,$ km s$^{-1}$ kpc$^{-1}$ & $28.52\pm0.12$ & $28.41\pm0.20$ & $28.65\pm0.15$ & $29.2\pm2.4$ \\
  $\Omega^{'}_0,$ km s$^{-1}$ kpc$^{-2}$ &$-4.021\pm0.032$ &$-3.994\pm0.053$ &$-4.106\pm0.038$ & $-3.41\pm0.56$ \\
  $\Omega^{''}_0,$ km s$^{-1}$ kpc$^{-3}$ & $0.783\pm0.025$ & $0.721\pm0.038$ & $0.956\pm0.045$ &            --- \\
   $\sigma_0,$ km s$^{-1}$  &          7.3  &          8.6  &           5.9 &       10.1 \\
        $V_0,$ km s$^{-1}$  & $228.1\pm4.4$ & $227.3\pm4.6$ & $229.2\pm4.5$ & $234\pm19$ \\
     $N_\star$       &          3981 &          1106 &          2426 &        449 \\
  \hline
 $\sigma_1,$ km s$^{-1}$ & $9.93\pm0.62$ & $13.37\pm0.90$ & $8.18\pm0.87$ & $16.9\pm1.4$\\
 $\sigma_2,$ km s$^{-1}$ & $8.75\pm0.59$ & $10.91\pm1.41$ & $6.68\pm0.38$ & $11.0\pm0.8$\\
 $\sigma_3,$ km s$^{-1}$ & $6.80\pm0.22$ & $ 8.47\pm0.37$ & $6.18\pm0.28$ & $ 6.7\pm2.6$\\
 $L_1, B_1$ & $~47^\circ,$ $~4^\circ$ & $~32^\circ,$  $-2^\circ$ & $~57^\circ,$ ~$13^\circ$ & $~33^\circ,$ $~6^\circ$\\
 $L_2, B_2$ & $137^\circ,$ $~6^\circ$ & $122^\circ,$ ~$-2^\circ$ & $149^\circ,$ ~$~8^\circ$ & $129^\circ,$ $46^\circ$\\
 $L_3, B_3$ & $283^\circ,$ $83^\circ$ & $~73^\circ,$ ~$87^\circ$ & $268^\circ,$ ~$75^\circ$ & $298^\circ,$ $44^\circ$\\
  \hline
 \end{tabular}\end{center} \end{table}
 \begin{table}[t]
 \caption[]{\small
The kinematic parameters found from the stars of the ALL sample with relative trigonometric parallax errors less than 20\%.
 }
  \begin{center}  \label{t:02}  \small
  \begin{tabular}{|l|r|r|r|r|r|}\hline
       Parameters & All stars & $(M_G)_0\leq2^m $ &   $(M_G)_0>2^m$ \\\hline

       $U_\odot,$ km s$^{-1}$ & $6.91\pm0.15$ & $6.23\pm0.29$ & $7.21\pm0.16$ \\
       $V_\odot,$ km s$^{-1}$ & $9.19\pm0.24$ & $8.96\pm0.48$ & $9.72\pm0.30$ \\
       $W_\odot,$ km s$^{-1}$ & $7.49\pm0.12$ & $7.01\pm0.21$ & $7.81\pm0.13$ \\

  $\Omega_0,$ km s$^{-1}$ kpc$^{-1}$ & $28.84\pm0.10$  & $28.64\pm0.15$  & $28.88\pm0.15$  \\
  $\Omega^{'}_0,$  km s$^{-1}$ kpc$^{-2}$ &$-4.063\pm0.029$ &$-4.035\pm0.044$ &$-4.114\pm0.040$ \\
  $\Omega^{''}_0,$ km s$^{-1}$ kpc$^{-3}$ & $0.766\pm0.025$ & $0.749\pm0.038$ & $0.919\pm0.048$ \\
   $\sigma_0,$ km s$^{-1}$  &          7.8  &          9.3  &           6.5 \\
        $V_0,$ km s$^{-1}$  & $230.7\pm4.4$ & $229.1\pm4.5$ & $231.0\pm4.5$ \\
     $N_\star$       &          4564 &          1995 &          2569 \\
  \hline
 $\sigma_1,$ km s$^{-1}$ & $11.22\pm0.75$ & $14.09\pm1.20$ & $8.77\pm0.83$ \\
 $\sigma_2,$ km s$^{-1}$ & $10.33\pm0.83$ & $11.31\pm1.71$ & $7.14\pm0.36$ \\
 $\sigma_3,$ km s$^{-1}$ & $ 7.42\pm0.24$ & $ 8.01\pm0.42$ & $6.38\pm0.28$ \\
 $L_1, B_1$ & $~66^\circ,$ ~$~0^\circ$ & $~50^\circ,$ ~$~0^\circ$ & $~66^\circ,$ ~$~2^\circ$ \\
 $L_2, B_2$ & $156^\circ,$ ~$~1^\circ$ & $140^\circ,$ ~$~1^\circ$ & $156^\circ,$ ~$-2^\circ$ \\
 $L_3, B_3$ & $343^\circ,$ ~$89^\circ$ & $121^\circ,$ ~$89^\circ$ & $194^\circ,$ ~$88^\circ$ \\
  \hline
 \end{tabular}\end{center} \end{table}

\begin{figure}[t]
{\begin{center}
   \includegraphics[width=0.85\textwidth]{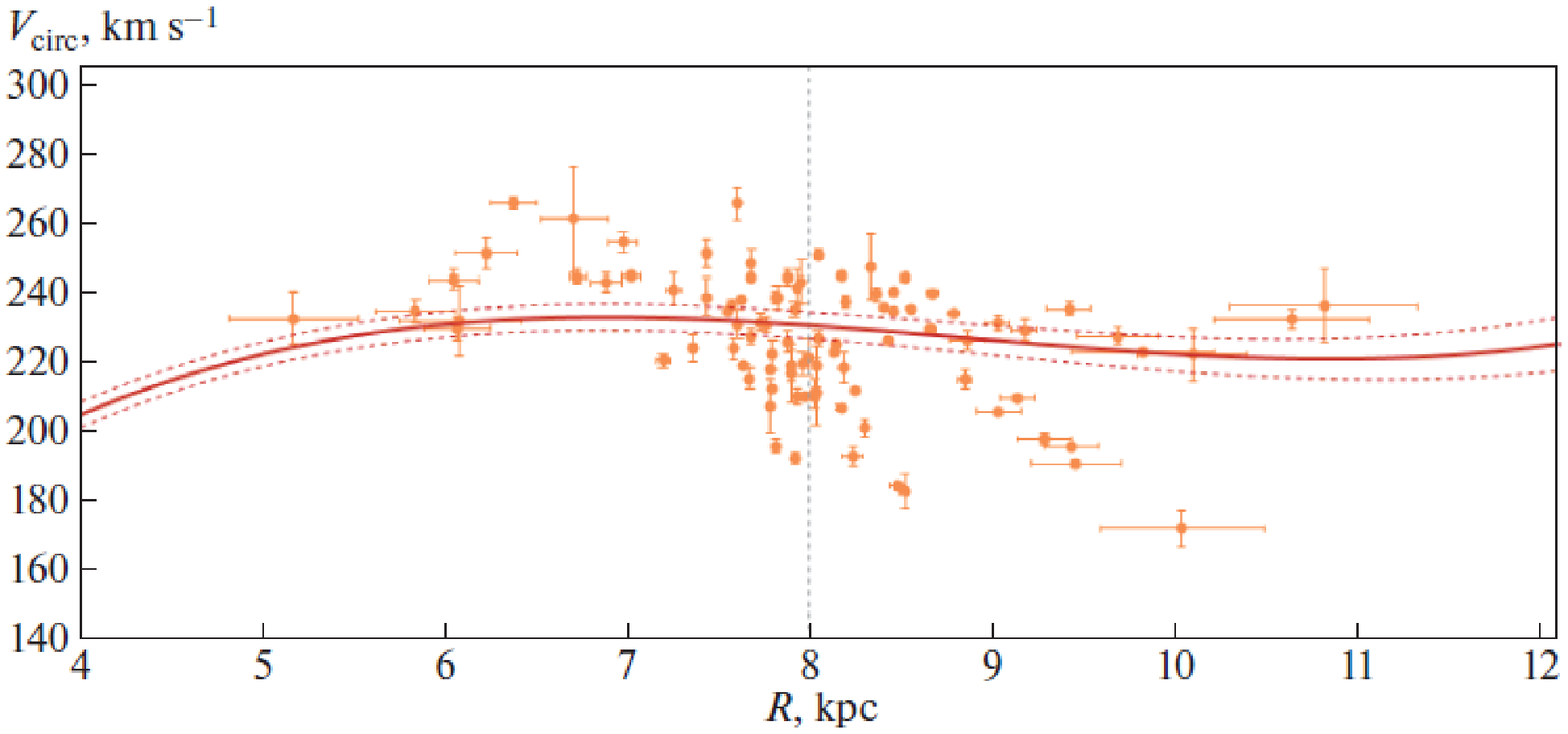}
 \caption{
The Galactic rotation curve constructed from the stars of the ALL sample with measured line-of-sight velocities. The vertical dashed line marks the Sun's position; the dashed curves indicate the boundaries of the $1\sigma$ confidence region.
   } \label{f-3}
\end{center}}
\end{figure}

Based on the stars of the PMS sample, we also found the kinematic parameters using the distances
calculated in [27]. With this approach we obtained
$(U_\odot,V_\odot,W_\odot)=(7.18,10.02,7.73)\pm(0.16,0.25,0.12)$~km s$^{-1}$ and the following parameters of the angular velocity of Galactic rotation:
 \begin{equation}
 \label{sol-55}
 \begin{array}{lll}
  \Omega_0=~28.59\pm0.13~\hbox{km s$^{-1}$ kpc$^{-1}$},\\
  \Omega^{\prime}_0=-4.047\pm0.035~\hbox{km s$^{-1}$ kpc$^{-2}$},\\
  \Omega^{\prime\prime}_0=~0.782\pm0.026~\hbox{km s$^{-1}$ kpc$^{-3}$}.
 \end{array}
 \end{equation}
The error per unit weight here is $\sigma_0=7.3$~km s$^{-1}$ and the linear circular rotation velocity of the solar neighborhood is $V_0=228.7\pm4.4$~km s$^{-1}$; we used 3612 stars. A comparison of the parameters (15) with those from the first column in Table 1 shows no significant differences between the two approaches. Below
we use the distances calculated via the original trigonometric parallaxes from the Gaia DSR2 catalogue, as
described in Section 3.3.

The kinematic parameters determined from the stars of the ALL sample are given in Table 2. We took
the stars with relative trigonometric parallax errors less than 20\% and heliocentric distances in the range of
distances from 0.5 to 4 kpc. The first, second, and third columns in the table give the solutions obtained
from all stars of the sample, the stars with $(M_G)_0\leq2^m$, and the stars with $(M_G)_0>2^m$, respectively. The parameters of the stellar residual velocity ellipsoid are given in the lower part of Table 2.

With the known parallaxes, proper motions, and line-of-sight velocities, the rectangular stellar space
velocities $U,V,W$ are calculated as follows:
\begin{equation}\label{UVW}
 \begin{array}{lll}
U=V_r \cos l \cos b-V_l \sin l-V_b \cos l \sin b,\\
V=V_r \sin l \cos b+V_l \cos l-V_b \sin l \sin b,\\
W=V_r \sin b+V_b \cos b.
\end{array}
\end{equation}
The circular linear velocity $V_{circ}$ directed along the Galactic rotation is then expressed as
  $$V_{circ}=U\sin\theta +(V_0+V)\cos\theta,$$
the angle $\theta$ satisfies the relation $\tan\theta=-Y/X,$ where $X$ and $Y$ are the Galactocentric rectangular coordinates of the star. Figure 3 shows the circular velocities as a function of for 96 stars with measured line-of-sight velocities. The Galactic rotation curve was constructed with the parameters listed in the first column of Table 2.

Krisanova et al. [11] studied the sample of young pre-main-sequence stars selected in [6]. The components of the group velocity vector $(U_\odot,V_\odot,W_\odot)=(9.99,14.04,7.25)\pm(0.13,0.22,0.10)$~km s$^{-1}$ and the following components of the angular velocity of Galactic rotation were found using the proper motions of
more than 25000 stars: 
 $\Omega_0 =28.40\pm0.11$~km s$^{-1}$ kpc$^{-1}$,
 $\Omega^{'}_0=-3.933\pm0.033$~km s$^{-1}$ kpc$^{-2}$, and 
 $\Omega^{''}_0=0.804\pm0.040$~km s$^{-1}$ kpc$^{-3}$, where the error per unit weight is $\sigma_0=16.0$~km s$^{-1}$ and the circular rotation velocity of the solar neighborhood around the Galactic center is
$V_0=227\pm4$~km s$^{-1}$ (for the adopted $R_0=8.0\pm0.15$~kpc). The stars within 3 kpc of the Sun with relative Gaia DR2 trigonometric parallax errors less than 10\% were taken.

We see that having a factor of 5 smaller number of stars, we obtain (the first column in Table 2) very close
values of the kinematic parameters even with their smaller errors. The Galactic rotation parameters
found in this paper are in excellent agreement with the results of an analysis of the youngest stars.

For example, based on OB stars with relative parallax errors less than 30\%, Bobylev and Bajkova [33]
determined the following parameters of the Galactic
rotation curve:
  $(U,V,W)_\odot=(8.16,11.19,8.55)\pm(0.48,0.56,0.48)$~km s$^{-1}$, 
  $\Omega_0=28.92\pm0.39$~km s$^{-1}$ kpc$^{-1}$,
  $\Omega^{'}_0=-4.087\pm0.083$~km s$^{-1}$ kpc$^{-2}$, and 
  $\Omega^{''}_0=0.703\pm0.067$~km s$^{-1}$ kpc$^{-3}$, 
where the circular velocity of the local standard of rest is $V_0=231\pm5$~km s$^{-1}$ (for the
adopted $R_0=8.0\pm0.15$~kpc).

Based on a sample of 147 masers, Reid et al. [34] found the following values of the two most important
kinematic parameters: $R_0=8.15\pm0.15$~kpc and $\Omega_\odot=30.32\pm0.27$~km s$^{-1}$ kpc$^{-1}$, where $\Omega_\odot=\Omega_0+V_\odot/R,$ and the velocity $V_\odot=12.24$~km s$^{-1}$ was taken from [35].
These authors used a series expansion of the linear Galactic rotation velocity. Based on a similar
approach, Hirota et al. [20] obtained the following estimates by analyzing 99 masers observed as part of
the VERA program: $R_0=7.92\pm0.16$\,(stat.)$\pm0.3$\,(syst.)~kpc and $\Omega_\odot=30.17\pm0.27$\,(stat.)$\pm0.3$\,(syst.)~km s$^{-1}$ kpc$^{-1}$, where $\Omega_\odot=\Omega_0+V_\odot/R,$ and the velocity $V_\odot=12.24$~km s$^{-1}$ was also taken from [35].

Bobylev [5] performed a kinematic analysis of the T Tauri stars from the list by Zari et al. [4]. For example,
when solving the kinematic equations using only the proper motions of the stars from the PMS3 sample, the
error per unit weight was $\sigma_0=6.9$~km s$^{-1}$. The following parameters of the residual velocity ellipsoid for the PMS3 sample stars were also determined in his paper:
$\sigma_{1,2,3}=(8.87,5.58,3.03)\pm(0.10,0.20,0.04)$~km s$^{-1}$.
The results obtained by determining the analogous characteristics from the significantly more distant
stars of the ALL sample (from the lower part of the HR diagram, $(M_G)_0>2^m$) are in excellent agreement with
the cited ones.

 \section{CONCLUSIONS}
In this paper we used young stars of various masses to determine the Galactic rotation parameters. These
stars were selected by Vioque et al. (2020) based on data from the Gaia DR2 catalogue by invoking the
photometric characteristics from such infrared surveys as 2MASS, WISE, IPHAS, and VPHAS+. The stars
under consideration are members of several samples. First, these include 693 candidates for classical Be
stars (CBe sample). Second, these include 1309 intermediate-mass pre-main-sequence stars (EITHER
sample). Third, these include a sample (PMS) of 8~470 candidates for young pre-main-sequence stars of various
masses.

Based on the proper motions of $\sim$4500 stars with relative trigonometric parallax errors less than 20\%
and heliocentric distances from 0.5 to 4 kpc, we found the following parameters of the angular velocity of
Galactic rotation: 
 $\Omega_0 =28.84\pm0.10$~km s$V_0=230.7\pm4.4$~ kpc$^{-1}$,
 $\Omega^{'}_0=-4.063\pm0.029$~km s$^{-1}$ kpc$^{-2}$, and
 $\Omega^{''}_0=0.766\pm0.020$~km s$^{-1}$ kpc$^{-3}$, where the Oort constants are
$A=16.25\pm0.33$~km s$^{-1}$ kpc$^{-1}$ and $B=-12.58\pm0.34$~km s$^{-1}$ kpc$^{-1}$ and the linear circular rotation velocity of the solar neighborhood around the Galactic center is $V_0=230.7\pm4.4$~km s$^{-1}$ (for the adopted $R_0=8.0\pm0.15$~kpc).

The residual velocity dispersion of the stars considered was shown to be low; this shows their youth. We
found that the residual velocity dispersion slightly depends on the stellar positions on the HR diagram.
For example, based on the stars from the lower part of the diagram with an absolute magnitude $(M_G)_0>2^m,$
we found the following parameters of their residual velocity ellipsoid: $\sigma_{1,2,3}=(8.77,7.14,6.38)\pm(0.83,0.36,0.28)$~km s$^{-1}$, while based on the stars from the upper part of the diagram with an absolute magnitude $(M_G)_0\leq2^m,$ we obtained
$\sigma_{1,2,3}=(14.09,11.31,8.01)\pm(1.20,1.71,0.42)$~km s$^{-1}$. Thus, the residual
velocity dispersion averaged over three coordinates is $\sim$11 km s$^{-1}$ for Herbig Ae/Be stars and $\sim$7 km s$^{-1}$ for T Tauri stars. The third axis of the ellipsoids found
from the ALL sample has no significant deflection from the vertical.

 \subsubsection*{ACKNOWLEDGEMENTS}
The authors thank the referee for useful remarks that helped us to improve the paper.

 \medskip\subsubsection*{REFERENCES}

 {\small
 \quad ~1. A. G. A. Brown, A. Vallenari, T. Prusti, de Bruijne, et al., Astron. Astrophys. 616, 1 (2018).

2. L. Lindegren, J. Hernez, A. Bombrun, and S. Klioner, et al., Astron. Astrophys. 616, 2 (2018).

3. T. Prusti, J. H. J. de Bruijne, A. G. A. Brown, A. Vallenari, et al., Astron. Astrophys. 595, 1 (2016).

4. E. Zari, H. Hashemi, A. G. A. Brown, K. Jardine, and P. T. de Zeeuw, Astron. Astrophys. 620, 172 (2018).

5. V. V. Bobylev, Astron. Lett. 46, 131 (2020).

6. G. Marton, P. Grah\'am, E. Szegedi-Elek, J. Varga, et al., Mon. Not. R. Astron. Soc. 487, 2522 (2019).

7. R. M. Skrutskie, R. M. Cutri, R. Stiening, M. D. Weinberg, et al., Astron. J. 131, 1163 (2006).

8. E. L. Wright, P. R. M. Eisenhardt, A. K. Mainzer, M. E. Ressler, et al., Astron. J. 140, 1868 (2010).

9. R. M. Cutri, E. L. Wright, T. Conrow, J. Bauer, et al., VizieR On-line Data Catalog: II/311 (2012).
http://wise2.ipac.caltech.edu/docs/release/allsky/expsup/index.html.
 
 10. R. Adam, P. A. R. Ade, N. Aghanim, M. I. R. Alves, et al., Astron. Astrophys. 594, 10 (2016).
 
 11. O. I. Krisanova, V. V. Bobylev, and A. T. Bajkova, Astron. Lett. 46, 370 (2020).

 12. M. Vioque, R. D. Oudmaijer, M. Schreiner, I. Mendigutia, D. Baines, N. Mowlavi, and R. P\'erez-Martinez,
Astron. Astrophys. 638, 21 (2020).

 13. J. E. Drew, R. Greimel, M. J. Irwin, A. Aungwerojwit, et al., Mon. Not. R. Astron. Soc. 362, 753 (2005).

 14. G. Barentsen, H. J. Farnhill, J. E. Drew, E. A. Gonz\'alez-Solares, et al., Mon. Not. R. Astron. Soc. 444,
3230 (2014).

 15. J. E. Drew, E. Gonz\'alez-Solares, R. Greimel, M. J. Irwin, et al., Mon. Not. R. Astron. Soc. 440, 2036 (2014).

 16. V. V. Bobylev and A. T. Bajkova, Mon. Not. R. Astron. Soc. 437, 1549 (2014).

 17. J. P. Valle\'e, Astrophys. Space Sci. 362, 79 (2017).

 18. R. de Grijs and G. Bono, Astrophys. J. Suppl. 232, 22 (2017).

 19. T. Camarillo, M. Varun, M. Tyler, and R. Bharat, Publ. Astron. Soc. Pacif. 130, 4101 (2018).

 20. T. Hirota, T. Nagayama, M. Honma, Y. Adachi, et al., Publ. Astron. Soc. Jpn. 72, 50 (2020).

 21. R. Abuter, A. Amorim, N. Baub\"ock, J. P. Berger, et al., Astron. Astrophys. 625, L10 (2019).

 22. T. Do, A. Hees, A. Ghez, G. D. Martinez, et al., Science (Washington, DC, U.S.) 365, 664 (2019).

 23. F. Arenou, X. Luri, C. Babusiaux, C. Fabricius, et al., Astron. Astrophys. 616, 17 (2018).

 24. L. N. Yalyalieva, A. A. Chemel, E. V. Glushkova, A. K. Dambis, and A. D. Klinichev, Astrophys. Bull.
73, 335 (2018).

 25. A. G. Riess, S. Casertano, W. Yuan, L. Macri, et al., Astrophys. J. 861, 126 (2018).

 26. J. C. Zinn, M. H. Pinsonneault, D. Huber, and D. Stello, Astrophys. J. 878, 136 (2019).

 27. C. A. L. Bailer-Jones, J. Rybizki, M. Fouesneau, G. Mantelet, and R. Andrae, Astron. J. 156, 58 (2018).

 28. T. E. Lutz and D. H. Kelker, Publ. Astron. Soc. Pacif. 85, 573 (1973).

 29. K. F. Ogorodnikov, Dynamics of Stellar Systems, Ed. by A. Beer (Pergamon, Oxford, 1965).

 30. A. Bressan, P. Marigo, L. Girardi, et al., Mon. Not. R. Astron. Soc. 427, 127 (2012).

 31. J. Tang, A. Bressan, P. Rosenfield, A. Slemer, et al., Mon. Not. R. Astron. Soc. 445, 4287 (2014).

 32. Y. Chen, L. Girardi, A. Bressan, P. Marigo, M. Barbieri, and X. Kong, Mon. Not. R. Astron. Soc. 444, 2525
(2014).

 33. V. V. Bobylev and A. T. Bajkova, Astron. Lett. 44, 676 (2018).
 
 34. M. J. Reid, N. Dame, K. M. Menten, A. Brunthaler, et al., Astrophys. J. 885, 131 (2019).

 35. R. Sch\"onrich, J. Binney, and W. Dehnen, Mon. Not. R. Astron. Soc. 403, 1829 (2010).

 }
 \end{document}